\documentclass[twocolumn,showpacs,prb,superscriptaddress]{revtex4}
\newcommand{\figurewidth}{\columnwidth}
\usepackage{graphicx}
\newcommand{\av}{_{\rm av}}
\def\bea{\begin{eqnarray}}
\def\eea{\end{eqnarray}}

\def\d{\delta}
\def\p{\partial} 
\def\nn{\nonumber}

\def\l{\lambda}

\def\f{\frac}

\def\s{\sigma}
\def\bd{\bar{\d}}

\begin{document}

\title{Ensemble dependence in the Random transverse-field Ising chain.}
\author{Abhishek Dhar}
\email{dabhi@bartok.ucsc.edu}
\affiliation{Physics Department, University of California, Santa Cruz,
CA 95064}
\affiliation{Raman Research Institute, Bangalore 560080}
\author{A. P. Young}
\email{peter@bartok.ucsc.edu}
\homepage{http://bartok.ucsc.edu/peter}
\affiliation{Physics Department, University of California, Santa Cruz,
CA 95064}
\date{\today} 

\begin{abstract}
In a disordered system one can either consider a microcanonical
ensemble, where there is a precise constraint on the random variables,
or a canonical ensemble where the variables are chosen according to a
distribution without constraints.
We address the question as to whether critical exponents
in these two cases can differ through a detailed study of the 
random transverse-field Ising chain. We find that the exponents are the
same in both ensembles, though some critical amplitudes vanish in the
microcanonical ensemble for correlations
which span the whole system and are particularly sensitive to
the constraint. This can \textit{appear} as a different exponent.
We expect that this apparent dependence of exponents on ensemble is related to
the integrability of the model, and would not occur in non-integrable models.
\end{abstract}

\maketitle

%\narrowtext

\section{Introduction}
\label{sec:intro}

In the study of the critical behavior of disordered systems, it is usual to
pick the random variables from some distribution. This allows
sample-to-sample fluctuations in
the sum of the interactions (e.g.~nearest neighbor)
of order
$\sqrt{N}$. We will call this the canonical
ensemble of disorder, by analogy with the canonical ensemble of
statistical mechanics which allows fluctuations in the energy. It is sometimes
of interest to complete this analogy and define a microcanonical ensemble
of the disorder in which there is a strict constraint, for
example by
fixing exactly the sum of the (e.g. nearest neighbor) interactions in
each sample.
Our experience from conventional
statistical mechanics tells us that in the thermodynamic limit the
choice of ensembles does not matter but it is not very clear that this
is also true for random systems, especially for the case of quantum phase
transitions.

In this paper we will study the simplest disordered model with a
quantum phase transition, the random transverse-field Ising chain (RTFIC)
\cite{fisher:95,fisher:98} with the Hamiltonian:
\begin{equation}
{\cal H} = -\sum_{i=1}^{L-1}  J_i \s^z_i \s^z_{i+1} -\sum_{i=1}^L h_i \s^x_i.
\end{equation}
where $J_i>0$ and $h_i>0$ are random variables chosen from distributions
$\rho(h)$ and $\pi (J)$ with averages $[\ln h]\av,~[\ln J]\av$
and variances ${\rm var}(\ln h),~{\rm var}(\ln J)$.  We use \textit{free}
boundary conditions, so the sum for the $J_i$ stops at $L-1$.

Let us define the two ensembles, microcanonical and canonical,
precisely for this model. For the canonical ensemble the $h_i$ and $J_i$ are
chosen randomly. A parameter which characterizes the deviation from criticality
is $\bd$
%\equiv [\delta]\av
where
\begin{equation}
\bd=\f{[\ln h]\av - [\ln J]\av} {{\rm var}(\ln h)+ {\rm var}(\ln J)} ,
\label{bd}
\end{equation}
For the microcanonical ensemble we 
constrain the
$h_i$ and $J_i$ such that the parameter
\begin{equation}
\delta = \f{1}{L-1} \f{%\displaystyle
\sum_{i=1}^L \ln h_i - \sum_{i=1}^{L-1} \ln J_i
-[\ln h]\av }
{{\rm var}(\ln h) + {\rm var}(\ln J)}
\label{delta}
\end{equation}
is set to a prescribed value for \textit{each} sample.
The last term in the numerator, which is not necessary to get the asymptotic behavior, corrects
for there being one more $h_i$ than $J_i$ with free boundary conditions.
It ensures that $\bd =  [\delta]\av$ even for a finite system.
%\begin{equation}
%\bd=\f{[\ln h]\av - [\ln J]\av} {{\rm var}(\ln h)+ {\rm var}(\ln J)} ,
%\label{bd}
%\end{equation}
%is specified.
In the canonical ensemble, the
fluctuations in $\delta$ from sample to sample are 
$O(1/\sqrt{L})$. It is known that a phase transition occurs in this
model at
$\bd=0$. All our numerical results in the paper will be at the critical point.

Pazmandi et al.\cite{pazmandi:97}
have argued that it is precisely the $O(1/\sqrt{L})$ fluctuations
which lead to the 
bound on the finite size correlation length
exponent\cite{chayes:86,chayes:89,harris:74}
$\nu \geq 2/d$. Further, 
they argue that exponents in the microcanonical ensemble need
not satisfy this bound. In addition, Igloi and Rieger\cite{igloi:98} claim that
the exponents of the RTFIC can depend on the choice of ensemble. In this
paper
our main goal is to investigate these claims through a detailed study of the
zero-temperature critical properties of the RTFIC.  

Using the Jordan-Wigner transformation, the Hamiltonian of the RTFIC
can be mapped to a free-fermion problem and this mapping is
particularly useful in the context of numerical computations. It can
be shown that various physical quantities can be expressed in
a straightforward  way in terms of the eigenvalues and eigenstates of
the free Hamiltonian, which are easy to evaluate numerically. 
These have been discussed by several authors in  earlier papers
\cite{young:96,igloi:98}
and we will not repeat the derivations here but will use those results
in our numerical calculations. In this paper we look at the surface
and bulk magnetizations, the end-to-end correlation function and the
energy gap.  
For the surface magnetization, the free fermion 
method leads to a simple form and it is possible to obtain some detailed
results analytically. We first discuss this in Sec.~\ref{sec:surfmag}
and then present
numerical results for various other quantities in Secs.~\ref{sec:bulkmag}
and \ref{sec:endtoend}.  Our conclusions are summarized in
Sec.~\ref{sec:discussion}.

\section{Surface Magnetization} 
\label{sec:surfmag}

The simplest quantity to calculate is the
surface magnetization, $m^s$, which is defined
%for an open chain, i.e. one
with free boundary conditions in which we fix $\s^z$ at one end, $i=L$ say,
to be $+1$. The surface magnetization is then 
the expectation value of $\s^z$ at the other
end ($i=1$). This is equivalent to deleting the transverse field on site $L$, so
$\s^z_L$ commutes with the Hamiltonian and the
ground state is \textit{exactly} doubly degenerate, and calculating the
expectation value of $\s^z_1$ in the ground state with $\s^z_L=1$. Let us
denote this state by $|\tilde{0}\rangle$ and so
\begin{equation}
m_s = \langle \tilde{0}|\s^z_1|\tilde{0} \rangle .
\end{equation}
This has a 
simple form\cite{peschel:84,igloi:98}, namely: 
\begin{equation}
m^s= \left[1+\sum_{i=1}^{L-1} \prod_{j=1}^{i}
\left({h_j \over J_j} \right)^2\right]^{-1/2}.
\label{smeq}
\end{equation}
Igloi and Rieger \cite{igloi:98} used this to numerically compute the
distributions of 
$B=-\log(m^s)$, $P_c(B)$ and $P_{mc}(B)$, in the two ensembles.

However it is also
possible to obtain the distribution functions analytically 
\cite{fisher:up} for $L \to \infty$
and we rederive those results here. 
First consider the canonical case. Let $x_l= (h_l/J_l)^2$. 
Then from Eq.~({\ref{smeq}) we get
\begin{eqnarray}
%B_L &=& \f{1}{2 } \ln{ [ 1+ x_1 (1+ x_2 +x_2 x_3+...x_2...x_{N-1})]}
B_L &=& \f{1}{2 } \ln \left[1\!+ x_1 (1\!+ x_2(1\!+ x_3(1\!+\!
\ldots (1+x_{L-1})\ldots)))\right]
\nn \\
&=& \f{1}{2} \ln [1 +x_1 e^{2 B_{L-1}}]   \nn \\
& \approx & \f{1}{2 } \ln x_1 + B_{L-1}   \, .
\label{rwlk}
\end{eqnarray} 
The above approximation is good most of the time since $B_L$ is
expected to be order $L^{1/2}$. For small $B_L$ we notice that the increment
in $B_L$ is always positive and so $B_L$ can never become negative.  
This and Eq.~(\ref{rwlk}) means that $B_L$ can be effectively
described by a biased random walk (in which $L$ is the time variable) with a
reflecting wall at the origin. 
It is convenient to introduce a scaled length variable 
\begin{equation}
\ell = L \f{{\rm var}(\ln h) + {\rm var}(\ln J)}{2} ,
\label{ell}
\end{equation}
in terms of which 
the probability distribution $P^c(B,L)$ can be
written as 
\begin{equation}
P^c(B,L) = \widetilde{P}(B, \ell).
\end{equation}
Then in the continuum limit, it is easy to see that $\widetilde{P}(B,
\ell)$ satisfies the following equation:
\begin{equation}
\f{\p \widetilde{P}}{\p \ell} = \f{\p^2 \widetilde{P}}{\p B^2} -
2 \bd \,\f{\p \widetilde{P}}{\p B} ,
\end{equation}
where
$\bd$ is given by Eq.~(\ref{bd}).
The reflecting boundary condition is imposed by requiring the current
at the origin $B=0$ to be zero, thus $ [\f{\p \tilde{P}}{\p B}-2 \bd \tilde{P}
]_{B=0}=0$. This problem is mathematically equivalent to Brownian
motion in a gravitational 
field and its solution, for identical boundary conditions, is
discussed in Ref.~\onlinecite{chandrasekhar:43}.
With the initial condition $\tilde{P}(B,
L=0)=\d (B)$ we find that the solution of the above equation is\cite{fisher:up}:
\begin{eqnarray}
& & P^{c}(B, L)= \widetilde{P}(B,\ell)=  \label{dist-c}
\\ 
& & \theta(B) \left[ \f{1}{(\ell \pi)^{1/2}}
e^{-(B-2 \ell \bd)^2/4\ell}-\bd\, e^{2
\bd B} {\rm erfc}\left(\f{B+2 \bd \ell}{2 \ell^{1/2}}\right) \right] . \nn
\end{eqnarray}
where $\rm {erfc}$ is the complementary error function.

For the microcanonical ensemble
the distribution $P^{mc}(B, L)$ can be found using the result that it
is related to $P^{c}(B, L)$ through the general
transformation Eq.~(\ref{reln}). One then gets\cite{fisher:up}:
\begin{equation}
P^{mc}(B, L)=2 \theta(B)\, \theta( B- 2 \d \ell) \left(\f{B}{\ell}-\d\right)
e^{-\f{B^2}{L}+2 B \d} .
\label{dist-mc}
\end{equation} 

Note that $P^{c}(B, L)$ is a function of the two scaling variables
$b = B/\ell^{1/2}$ and $\bd \ell^{1/2}$, and similarly $P^{mc}(B, L)$ is a
function of $\d \ell^{1/2}$ as well as $b$. According to finite size
scaling, the 
scaling variable associated with the deviation from criticality ($\d$ or $\bd$
here) should be proportional to $L^{1/\nu}$.
Hence Eqs.~(\ref{dist-c}) and (\ref{dist-mc}) show that the
true correlation length, as determined from finite size
scaling, is $\nu = 2$.

Now that we have the complete distributions for $B=-\ln(m^s)$ we can
calculate the \textit{mean} surface magnetization. Even though we find that 
$[ -\ln (m^s) ]\av \sim L^{1/2}$ in both ensembles, the behaviour of the
mean of $m^s$ (rather than its log) is quite different in the two ensembles.
This is because
$m^s=e^{-B}$ and so the main contribution to $m^s$ comes from
the behaviour of $P(B)$ at small $B$ (i.e. from \textit{rare} samples
with $m^s\approx 1$). 
For large $L$ we find the following asymptotic forms for the mean
magnetization. In the canonical case:
\begin{equation}
%\la m \ra_c= \f{e^{-\bd^2 L}}{(\pi L)^{1/2}}- \bd {\rm erfc}[\bd L^{1/2}] 
[m^s]\av^c= \f{e^{-\bd^2 \ell}}{(\pi \ell)^{1/2}}- \bd \,
{\rm erfc}[\bd \ell^{1/2}] 
\label{ms-c}
\end{equation}
while in the microcanonical ensemble we get: 
\begin{eqnarray}
[m^s]\av^{mc} &=& \f{2}{\ell}-2\d  \qquad \quad (\d <0) \nn \\ 
&=& e^{-2 \d \ell} (2 \d +\f{1}{\ell}) \quad (\d >0) .
\label{mc-mc}
\end{eqnarray}

For $\bd \ell^{1/2} \gg 1$, Eq.~(\ref{ms-c}) gives $[m^s]\av^c \sim
\exp(-\ell/\xi_c)$ where $\xi_c = 1/\bd^2$, in agreement with
$\nu = 2$ deduced earlier. However, for the microcanonical distribution with
$\d \ell \gg 1$,
we have $[m^s]\av^{mc} \sim \exp(-\ell/\xi_{mc})$ with $\xi_{mc} = 1/\d$. This
looks like an apparent correlation length exponent of 1 rather than 2.
However, it is worth investigating the origin of this discrepancy between the
apparent exponents in the two ensembles. For both ensembles the scaling
variable is $\d$ (or $\bd$)$\, L^{1/2}$. In the canonical ensemble, the
distribution in Eq.~(\ref{dist-c})
has a constant weight at $B=0$, which leads to the expected $\xi_c
= 1/\bd^2$. However, for the microcanonical ensemble, there is a ``hole'' in
the distribution for $B < 2 \d \ell$. The average is dominated by the
part of the distribution with the smallest $B$, so this difference in
the distributions for small $B$ accounts for the difference in the behavior of
the average.  Since the weight of the distribution for the
microcanonical ensemble vanishes at small $B$, we
argue that the \textit{amplitude} of the expected $1/\d^2$ divergence
of the correlation length for $[m^s]\av^{mc}$ is zero, and that the resulting
behavior, $\sim 1/\d$, is really a \textit{correction} to scaling. In the rest
of this paper we shall
reinforce the conclusion that $\nu=2$ in both ensembles
but with the leading amplitude vanishing, in the microcanonical case, for
certain quantities which are particularly sensitive to the microcanonical
constraint. This point of view is different from that of Igloi and
Rieger\cite{igloi:98} who argue that $\nu$ is different for the two ensembles.

For $L \to \infty$ and $\d$ (or $\bd$)$\, < 0$, we get
$[m_s]\av = -2\delta$ 
in both ensembles, giving a
magnetic exponent $\beta=1$.
At the critical point we find that the mean
magnetization decays with system size as: 
\begin{eqnarray}
[m^s]\av^c & \sim & \f{1}{L^{1/2}} \label{mdec}\\
{[m^s]}\av^{mc} & \sim & \f{1}{L} .\label{mcmdec}
\end{eqnarray}
From finite size scaling we expect a decay $\sim
{L^{-\beta/\nu}}$ where $\nu$ is the correlation length
exponent.
While this might suggest that $\nu=2$ in the canonical ensemble and $\nu=1$ in
the microcanonical case, we feel, as discussed above, that a more consistent
picture is that the amplitude of the leading divergence of the correlation
length appropriate to $[m^s]\av^{mc}$ vanishes for the microcanonical
ensemble and that the true exponent is $\nu=2$ in both cases.

\section{Bulk magnetization}
\label{sec:bulkmag}

In the previous section we saw that the correlation length exponent
for the surface magnetization seems, at first glance, to be
different in the canonical and microcanonical ensembles, but we argued that
the correct interpretation is that the exponents are the same, $\nu=2$,
but the amplitude of the expected divergence of certain quantities is
actually zero for the microcanonical ensemble. In this section
we strengthen this argument by
investigating the magnetization in the \textit{bulk} of the
sample, when a spin at the end is fixed.
We see the \textit{same} exponent $\nu =2$ in both ensembles, clearly
indicating that a correlation length with a $\nu=2$ divergence
does exist for the
microcanonical ensemble. Its absence in the surface
magnetization presumably indicates 
that the leading amplitude vanishes for this
quantity.

We again  consider an open chain with the spin at one end fixed to $\s^z_L=1$
and look at the magnetization at the middle of the chain
\begin{equation}
m=\langle \tilde{0}|\s^z_{L/2}|\tilde{0} \rangle . 
\label{bulk}
\end{equation}
Using the free fermion method this can be expressed as the determinant of a
matrix whose elements are expressed in terms of eigenstates of a quadratic
Hamiltonian. We evaluate this numerically and compute both the mean of the bulk
magnetization and also its  distribution in the two different ensembles. Here
we will only examine the data at the critical point. 

Since the spin at one end is fixed,
there are equal numbers of $h_i$ and $J_i$, 
so the definition of $\delta$ in Eq.~(\ref{delta}) is slightly modified, which
leads to the condition
\begin{equation}
\sum_{i=1}^{L-1} \left( \ln h_i - \ln J_i \right) = 0
\label{micro_crit}
\end{equation}
for criticality (i.e. $\delta = 0$) in the microcanonical ensemble.
We set $J=1$ and allow $h$ to take
values $2$ and $1/2$. In the canonical case each $h_i$ takes one of
its two values with equal probability. In the microcanonical case {\it
exactly} half of the sites, chosen at random,
are assigned $h=2$ and the other half
are assigned $h=1/2$, which clearly satisfies Eq.~(\ref{micro_crit}).

\begin{figure}
\includegraphics[width=\figurewidth]{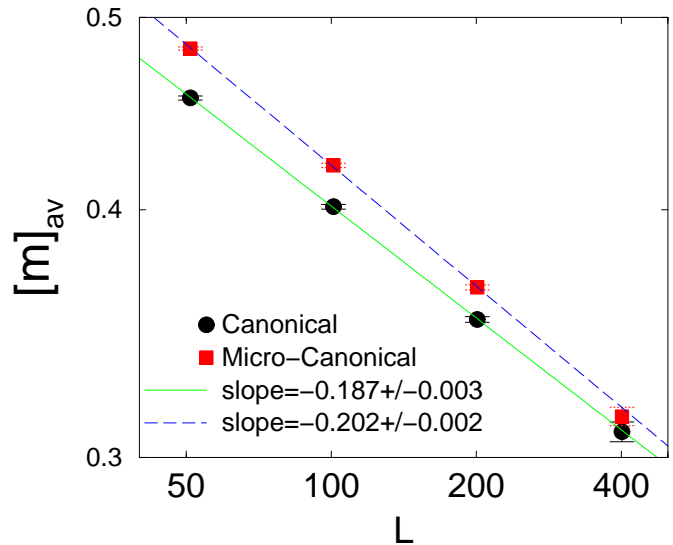}
\caption{ Mean of the bulk magnetization, as defined Eq.~(\ref{bulk}),
for different system
sizes, evaluated in the two  ensembles. The expected slope from
finite size scaling is $\beta/\nu \simeq 0.191$.}
\label{bmav}
\end{figure}

The numerical results for the decay of the mean
magnetization with system size are shown in Fig.~\ref{bmav}.
The mean is seen to behave similarly in
both ensembles and the system-size decay is consistent with the
form expected from finite-size scaling $[m]\av \sim L^{-\beta/\nu}$
with $\beta=(3-5^{1/2})/2$ and $\nu=2$, so that $\beta/\nu \simeq
0.191$.

We now look at the distribution of $m$. We use the variable
$b=-\ln (m)/L^{1/2}$ since this has good scaling properties. The
details of the distributions of $m$, shown in 
Figs.~\ref{bmdst1}, \ref{bmdst2}, and \ref{bmdst3}, are 
different in the two ensembles; in particular the microcanonical distribution
falls off faster at large argument.
However, as can be seen in
Fig.~\ref{bmdst3}, the behaviour at small values of the argument is
the same in both ensembles, which leads to the same asymptotic
behaviour for $[m ]\av$.

Thus, unlike the surface magnetization, the bulk magnetization shows the
\textit{same} critical behaviour in both ensembles. In particular, the results
of this section indicate that there is a correlation length which diverges
with exponent $\nu=2$ in the microcanonical ensemble. It is not seen in the
surface magnetization $m^s$, for which the correlation length diverges
less strongly 
with an exponent $\nu=1$, but this must simply indicate that the amplitude of
the $\nu=2$ divergence vanishes for $m^s$. 

%\vspace{-1.5cm}
\begin{figure}
\includegraphics[width=\figurewidth]{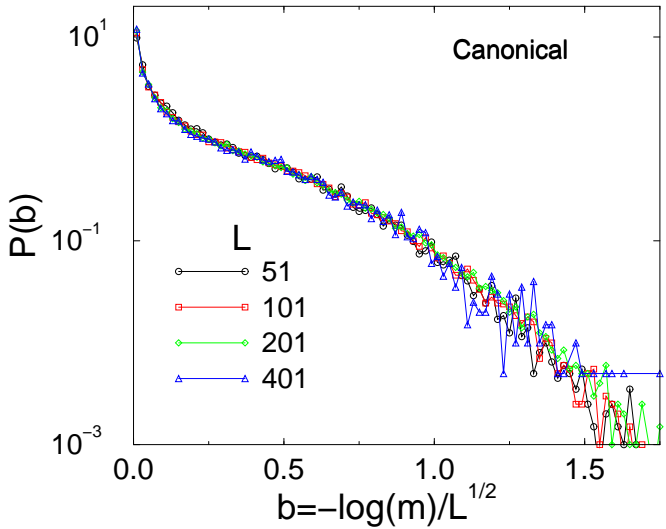}
\caption{ The distribution of the bulk-magnetization for different
system sizes for the canonical ensemble.  }
\label{bmdst1}
\end{figure}

%\vspace{-1.5cm}
\begin{figure}
\includegraphics[width=\figurewidth]{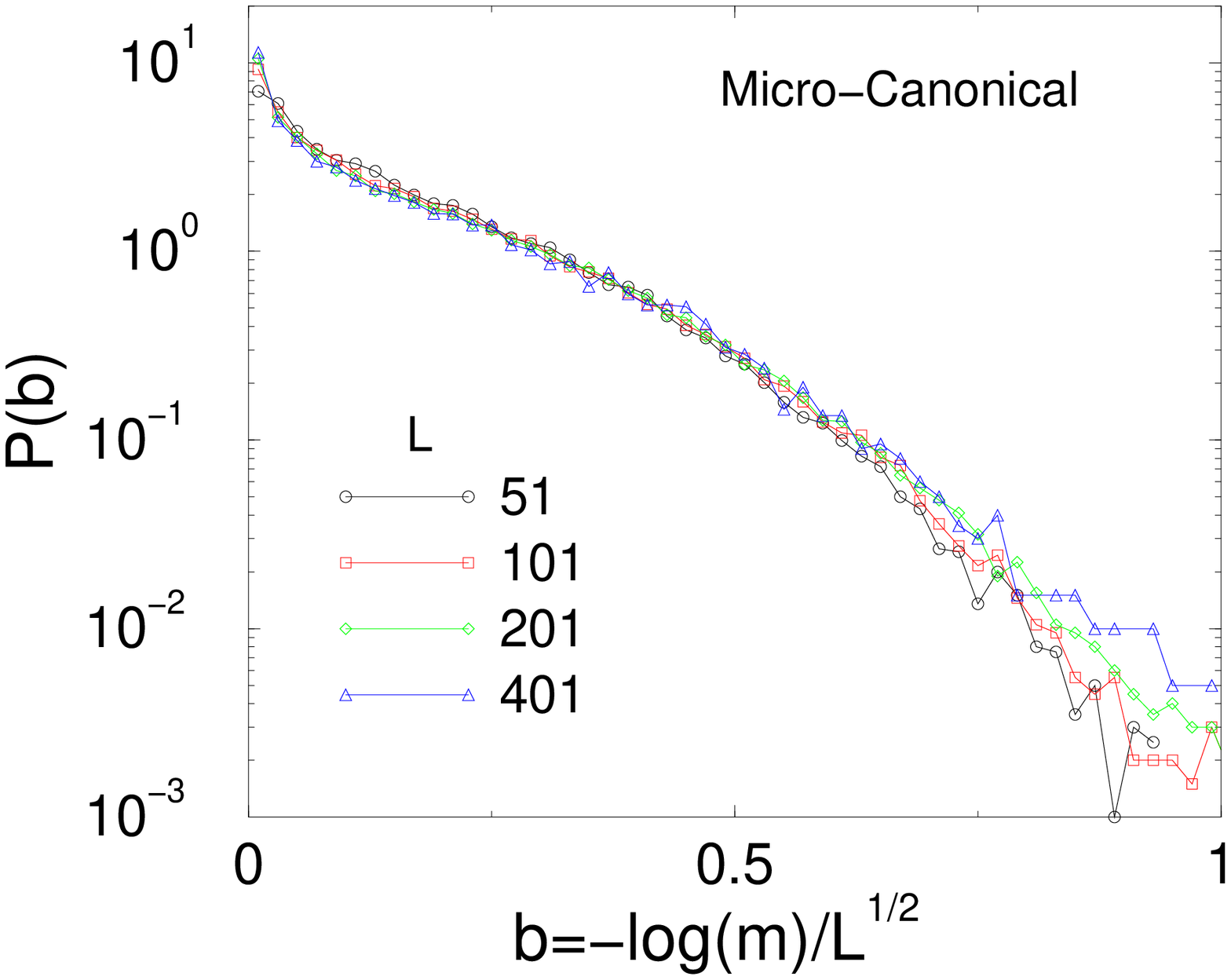}
\caption{ The distribution of the bulk-magnetization for different
system sizes for the microcanonical ensemble.}
\label{bmdst2}
\end{figure}

%\vspace{-1.5cm}
\begin{figure}
\includegraphics[width=\figurewidth]{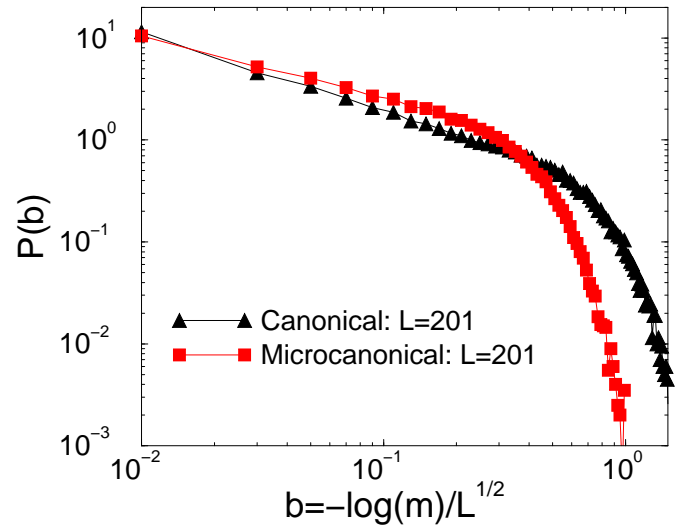}
\caption{ Comparison of the distributions of the bulk-magnetization
in the two ensembles for a chain of length $L=201$. On this
log-log plot they appear to have the same slope at small values of
the argument. However, at large argument, the microcanonical distribution
falls off faster.}
\label{bmdst3}
\end{figure}

\section{End-to-end correlations and gaps}
\label{sec:endtoend}

In this section we investigate numerically
the energy gap $\Delta$ and the end-to-end correlation
function
\begin{equation}
C_{1L}=\langle 0| \s^z_1 \s^z_L |0\rangle
\end{equation}
in the canonical and microcanonical ensembles to compare the
results of the two ensembles with each other and with
analytical results\cite{fisher:98} for the canonical ensemble. 

We take the following rectangular distribution for the
bonds and fields at the critical point
\begin{eqnarray}
\pi(J) & = & 
\left\{
\begin{array}{ll}
1 & \mbox{for $ 0 < J < 1$} \\
0  & \mbox{otherwise}
\end{array}
\right.
\nonumber \\ 
\rho(h) & = & 
\left\{
\begin{array}{ll}
1 & \mbox{for $ 0 < h < 1$} \\
0  & \mbox{otherwise.}
\end{array}
\right.
\label{dist}
\end{eqnarray}
which gives
\begin{eqnarray}
[\ln h]\av & = & -1, \qquad {\rm var}(\ln h) = 1 \nn \\
{[}\ln J]\av & = & -1, \qquad {\rm var}(\ln J) = 1 .
\label{meanvar}
\end{eqnarray}
From this it follows that
\begin{equation}
\ell = L ,
\end{equation}
where $\ell$ is defined in Eq.~(\ref{ell}).
We use free boundary conditions without constraining either
of the end spins.
From Eqs.~(\ref{delta}) and (\ref{meanvar})
the condition for criticality in the microcanonical
ensemble is
%This ensures that the scaled fields
%$h_i$ and bonds $J_i$ satisfy the microcanonical
%constraint  
\begin{equation}
\sum_{i=1}^L \ln h_i  -\sum_{i=1}^{L-1} \ln J_i= -1
\quad\mbox{(microcan.)} \, .
\label{constraint}
\end{equation}
We initially generate the $h_i$ and $J_i$
in an unconstrained way, as for the canonical ensemble, but then 
rescale $h_i$ by an appropriate factor so that the above condition is
satisfied. 

The distributions of $C_{1L}$ and $\Delta$
were studied earlier by Fisher and Young\cite{fisher:98} and we 
summarize some of their main results
for $L \to \infty$:  
\begin{eqnarray}
[ C_{1L}]\av^c & \sim & \f{1}{L}
\label{C1L}\\
{[} \Delta]\av^c & \sim & L^{1/6}
\exp\left[-\f{3}{2}\left(\f{\pi^2 L}{2}\right)^{1/3}\right]
\label{deltaav} \\
\ln(\Delta)-\ln(C_{1L}) & = & \sum_{i=1}^L \ln (h_l)-\sum_{i=1}^{L-1}\ln (J_l).
\label{fyan}  
\end{eqnarray}

\begin{figure}
\includegraphics[width=\figurewidth]{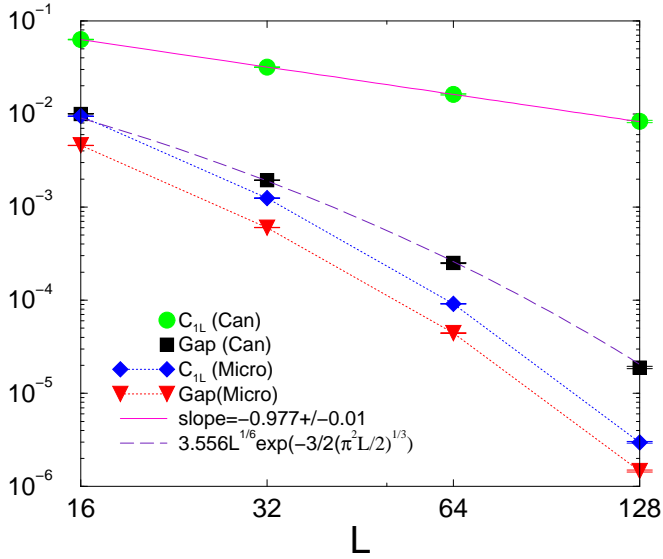}
\caption{Plot of the mean values of the end-to-end correlation
function $C_{1,L}$ and the energy gap $\Delta$ for both the
canonical and microcanonical ensembles. In the canonical case, the
fits are close to those predicted analytically and given in
Eqs.~(\ref{C1L}) and (\ref{deltaav}). For microcanonical ensemble,
the lines are
just guides to the eye.}
\label{avercg}
\end{figure}

In Fig.~\ref{avercg} we compare the system size dependence of the
average correlation function and the gap in the two ensembles.
For the canonical case, the data agree well with the analytic
predictions in Eqs.~(\ref{C1L}) and (\ref{deltaav}), as was also found in
Ref.~\onlinecite{fisher:98}. However, if we fit the data for the gap to
Eq.~(\ref{deltaav}) adjusting only the overall amplitude the $\chi^2$ is
150 which is very high. Hence there must be systematic corrections to
Eq.~(\ref{deltaav}) which are larger for the sizes studied than the, very
small,
statistical errors.

For the microcanonical ensemble, the data for
\textit{both} $[C_{1L}]\av^{mc}$ and
$[\Delta]\av^{mc}$ in Fig.~\ref{avercg} appear to decay as stretched
exponentials, and we
will discuss fits to this data below.

%In Fig.~\ref{mcCdec} we try to fit the data to the form
%$y=ae^{-bx^c}$ with $a,b$ and $c$ as fitting parameters. We take trial
%values of the parameter $c$ and do a linear best  fit with the
%variables $Y=\ln (y)$ and $X=x^c$. The best value of $c$ is then
%obtained as the one which gives a minimum $\chi^2$ value (see inset in
%Fig.~\ref{mcCdec}). This gives a value close to $3/8$ which is much
%higher than $1/3$ for the canonical case (note that visually both of the two
%powers seem to fit the data quite well). 
%However because of finite size effects and also because there may be
%power law corrections to the exponential decay form, it is difficult
%to arrive at any definite conclusions on the power. Later in this
%section we will give some further evidence which suggests that the
%power $c$ could be different from $1/3$.   

%\begin{figure}
%\includegraphics[width=\figurewidth]{fig6.eps}
%\caption{ 
%Plot of $[C_{1L}]av^{mc}$ for different system sizes, for the
%microcanonical ensemble. Best fits to stretched exponentials
%$ae^{-bx^c}$ with  powers $c=0.375$ (with $\chi^2=1.55$) and $c=1/3$
%($\chi^2=84.6$) are  also shown. The inset plots $\chi^2$ against $c$.}
%\label{mcCdec}
%\end{figure}
%

\begin{figure}
\includegraphics[width=\figurewidth]{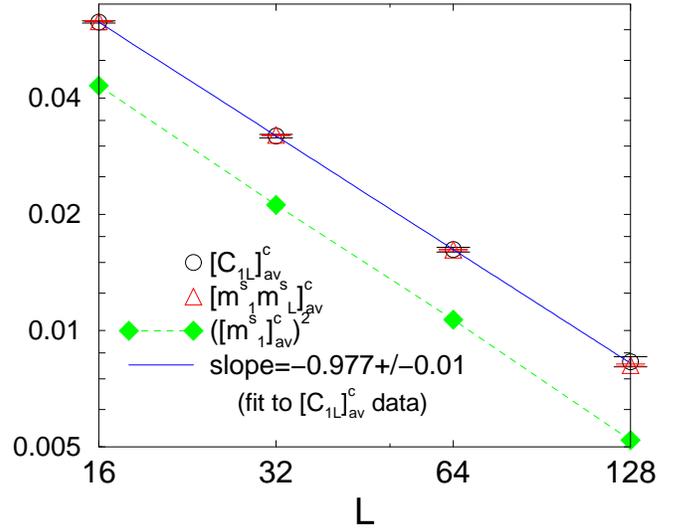}
\caption{ Plot of the mean correlation function $[C_{1,L}]\av^c$ versus
system size compared with $[m_1^s m_L^s]\av^c$ and
$([m_1^s]\av^c)^2$, for the canonical
ensemble. The  $([m_1^s]\av^c)^2$ data falls as $1/L$ as expected
from Eq.~(\ref{mdec}). 
\label{mmav}}
\end{figure}

\begin{figure}
\includegraphics[width=\figurewidth]{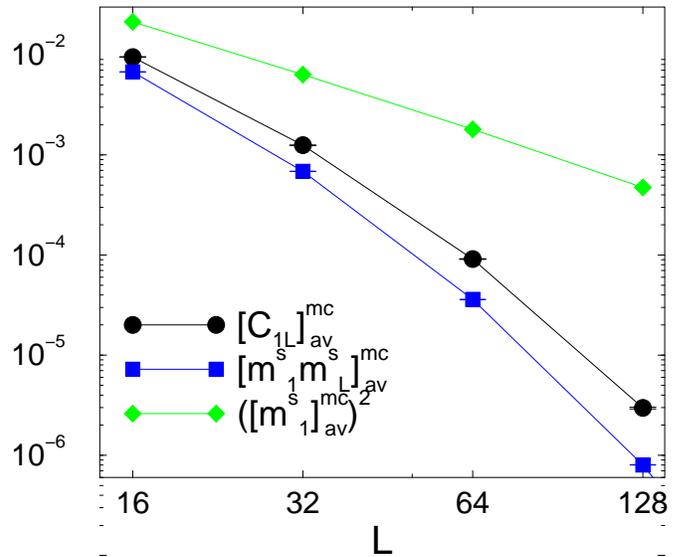}
\caption{Plot of the mean correlation function $[C_{1,L}]\av^{mc}$ versus
system size compared with $[m_1^s m_L^s]\av^{mc}$ and $([m_1^s]\av^{mc})^2$,
for the microcanonical
ensemble. 
The data for $([m_1^s]\av^{mc})^2$ decays as $1/L^2$ as expected from
Eq.~(\ref{mcmdec}). 
\label{mcmmav}}
\end{figure}

Interestingly, the value of $[C_{1L}]\av$
is  found to be close to $[m_1^s m_L^s]\av$,
where $m^s_1$ ($m^s_L$) is the surface magnetization at site
$1$ ($L$) with the spin at site $L$ ($1$) fixed. This can be seen in
Figs.~\ref{mmav} and \ref{mcmmav} where we plot both these quantities, as well
as  $[ m_1^s]\av^2$, for
different system sizes. Especially in the canonical case, $[C_{1L}]\av$ and
$[m_1^s m_L^s]\av$ are
almost indistinguishable. Note that for
the microcanonical, but not the canonical, ensemble $[ m_1^s]\av^2$
is much greater than $[m_1^s m_L^s]\av$. The reason for this is that
$[m_1^s]\av$ is
dominated by a few rare samples where the bonds are bigger than
the fields at the free end (and hence, for the microcanonical ensemble,
must be less than the fields at 
the other end because of the constraint). Hence for these
samples $m_L^s$ is smaller than typical in the microcanonical ensemble.

Since
$[C_{1L}]\av$ and $[ m_1^s m_L^s ]\av$ behave similarly we can
obtain a better
estimate for the decay law in the microcanonical case. This is
because $m_1^s m_L^s$ can be obtained directly from Eq.~(\ref{smeq})
and thus be accurately
computed numerically for bigger systems.
%The data is shown in
%Fig.~\ref{mmdec} where we find  a best fit to a stretched
%exponential with a power $c=0.366$. This is again much higher than the
%value $c=1/3$ (a fit with this power is quite bad, even visually, as
%can be seen in Fig.~\ref{mmdec}).
We assume the same form as in the exact result for the gap in the canonical
case, Eq.~(\ref{deltaav}), i.e.  $a L^p \exp(-bL^{\mu})$, and take $\mu
= 1/3$ the same value as in Eq.~(\ref{deltaav}).
The data  shown in 
Fig.~\ref{mmdec2} is fitted to 
$a L^p \exp(-bL^{1/3})$ by varying $a$ and $b$ for several (fixed) values
of $p$.
The minimum $\chi^2$ of 3.9, which
is quite acceptable for three degrees of freedom, is
obtained for $p\simeq0.44$. It we assume that $p =1/6$, as for the
canonical case, then $\chi^2 = 1380$ which is extremely high. However, we
noted for the canonical case, that there appear to be corrections to the
scaling form in Eq.~(\ref{deltaav}). Hence we cannot rule out the possibility
that $p=1/6$ also for the microcanonical ensemble.
%*** If we fix $p=1/6$ and vary
%$\mu$ what do we get for $\mu$? It might be interesting to give this to
%indicate the range of possible values of $\mu$.
%***
%In Fig.~\ref{mmdec2} we see that this fits
%the data quite well.

From Figs.~\ref{avercg} and \ref{mcmmav}, it seems plausible that
$[ m_1^s m_L^s ]\av^{\rm mc},~ [C_{1L}]\av^{\rm mc}$ and $[\Delta]\av^{\rm mc}$
all vary in the same way in the microcanonical ensemble. If this is so then
the data for the gap 
is consistent with the stretched
exponential form $a L^{1/6} \exp(-b L^{1/3})$, for both canonical and
microcanonical ensembles, though there are some systematic corrections to this
for the range of sizes that can be studied. This is known to be \textit{exact}
for the canonical ensemble, see Eq.~(\ref{deltaav}).
%
%*** Rephrase the next 2 sentences ***
%
%If we assume this is true, then, from the fits to
%the data for $[ m_1^s m_L^s ]\av^{\rm mc}$, we find for the gap that
%that the power of $L$ in the exponent of the stretched
%exponential $(\mu)$ could be the same $(1/3)$ for both canonical and
%microcanonical ensembles. However, this requires that the power of $L$ in the
%prefactor is different in the two cases: $p=1/3$ (canonical) versus $p \simeq
%0.44$ (microcanonical).
It would be interesting to see if the dependence of gap on system size could
be determined analytically for the microcanonical ensemble using random walk
arguments.

\begin{figure}
\includegraphics[width=\figurewidth]{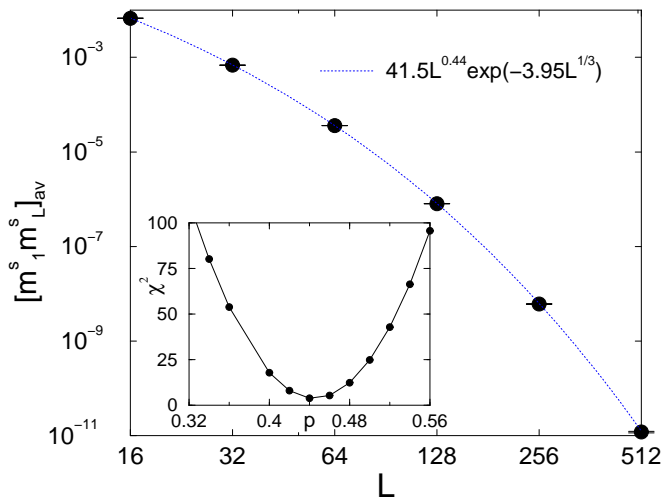}
\caption{ 
Plot of $[m_1^s m_L^s]\av^{mc}$ against $L$
for different system sizes, for the
microcanonical ensemble. A best fit to the form $a L^p e^{-bL^{1/3}}$
with $p=0.44$ ($\chi^2=3.9$) is shown. The inset plots $\chi^2$ against $p$.}
\label{mmdec2}
\end{figure}

\begin{figure}
\includegraphics[width=\figurewidth]{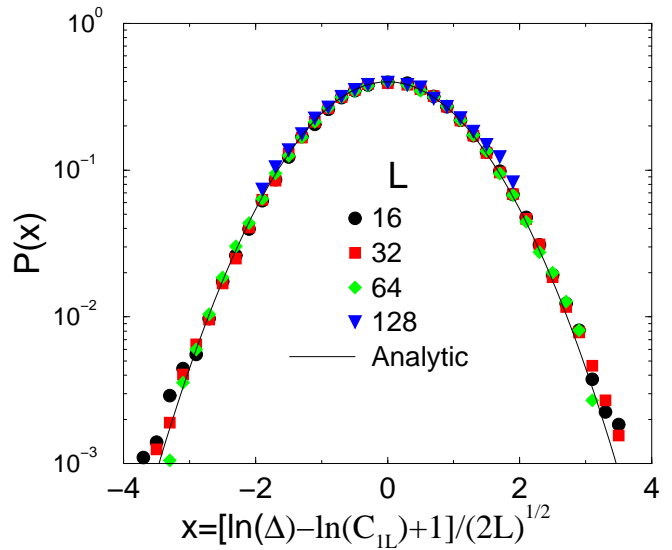}
\caption{Distribution of $[\ln(\Delta)-\ln(C_{1L})+1]/(2L)^{1/2}$ for
the canonical case. The analytic form, deduced from Eq.~(\ref{fyan}),
is a Gaussian with variance unity. This is shown by the solid line.
}
\label{diff}
\end{figure}

\begin{figure}
\includegraphics[width=\figurewidth]{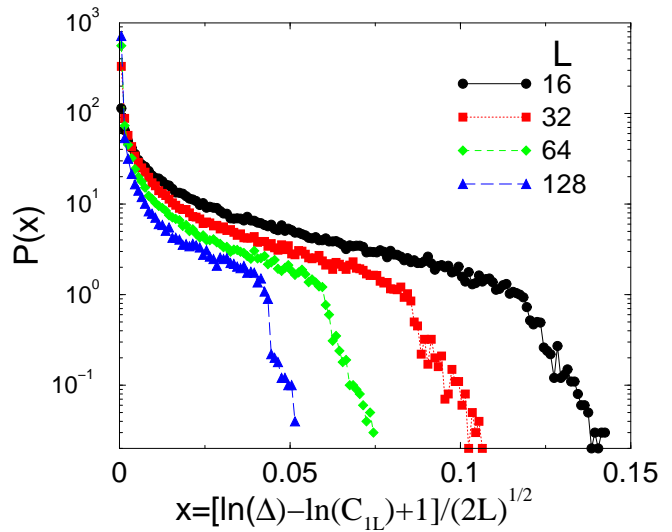}
\caption{Distribution of $[\ln(\Delta)-\ln(C_{1L})+1]/(2L)^{1/2}$ for
the microcanonical case. Equations (\ref{constraint}) and (\ref{fyan}) predict
that $\ln(\Delta)-\ln(C_{1L}) +1$ should be identically
zero in the thermodynamic limit. The data seems to be tending towards this for
large $L$.}
\label{mcdiff}
\end{figure}

\begin{figure}
\includegraphics[width=\figurewidth]{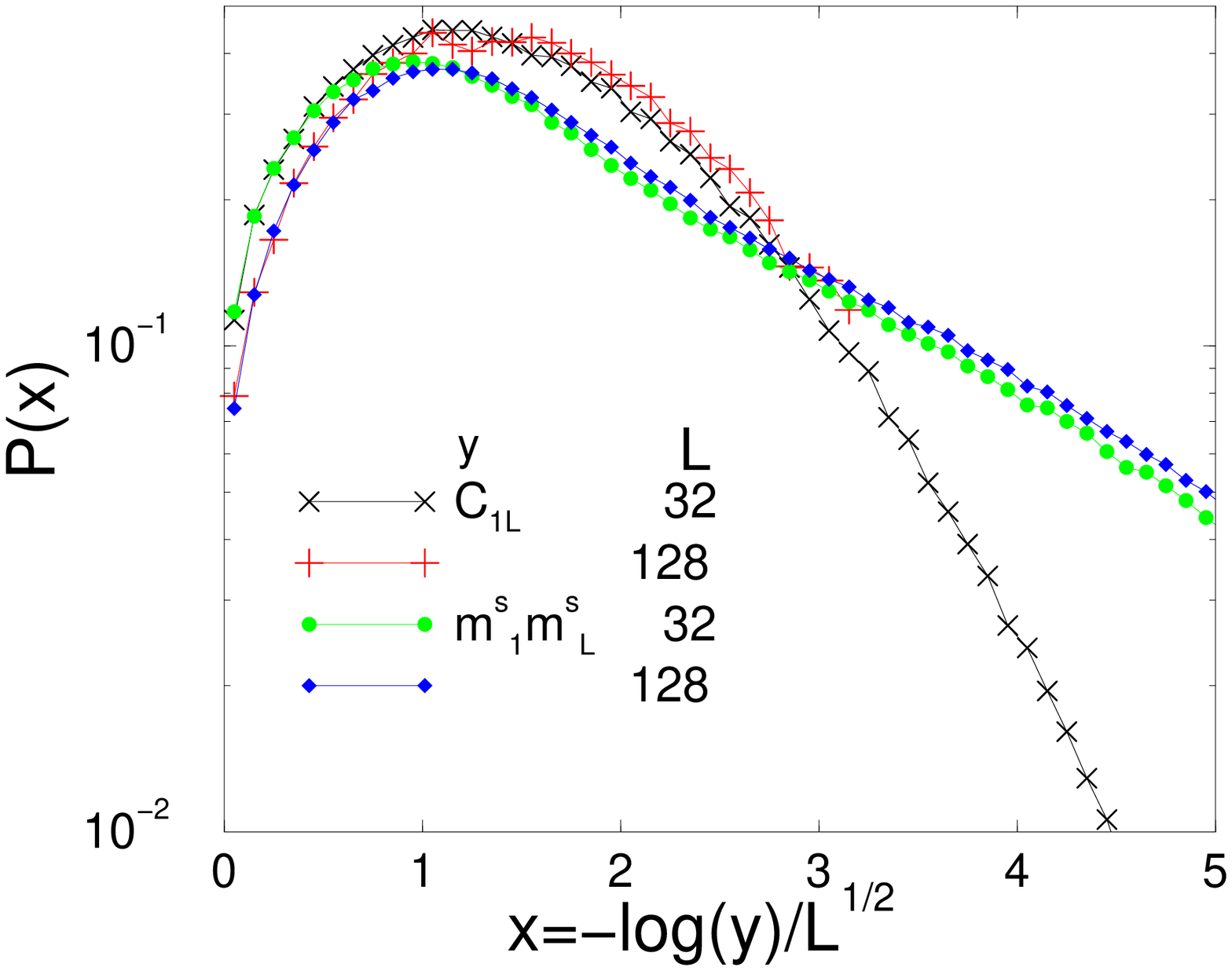}
\caption{Probability distributions of $C_{1L}$ and $m_1^s m_L^s$ for
system sizes $L=32$ and $128$ for the canonical ensemble. }
\label{cdist}
\end{figure} 

\begin{figure}
\includegraphics[width=\figurewidth]{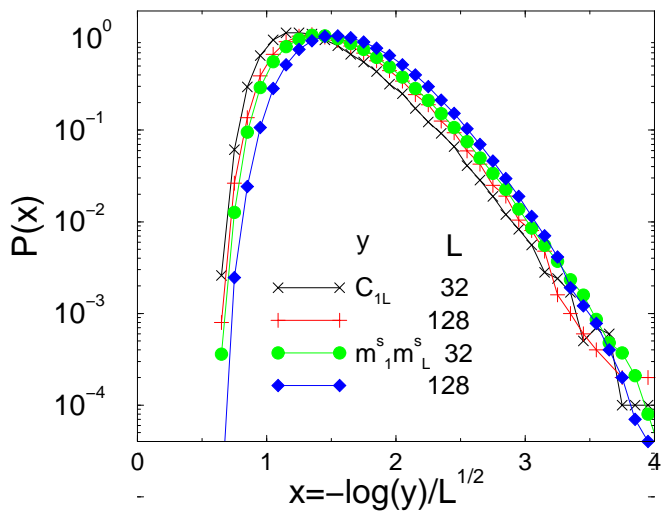}
\caption{Probability distributions of $C_{1L}$ and $m_1^s m_L^s$ for
system sizes $L=32$ and $128$ for the microcanonical ensemble. }
\label{mcdist}
\end{figure}

The distribution of the difference $\ln(\Delta)-\ln(C_{1L}) +1$
is plotted in 
Figs.~\ref{diff} and \ref{mcdiff}. 
In the canonical case, Eqs.~(\ref{fyan}) and (\ref{meanvar})
predict a Gaussian distribution with
zero mean and standard deviation $\sqrt{2L}$ for large $L$. 
Fig.~\ref{diff} shows that this works very well for the full range of sizes
studied numerically. Equations (\ref{constraint}) and (\ref{fyan}) predict
that $\ln(\Delta)-\ln(C_{1L}) +1$ should be identically
zero in the microcanonical
ensemble for $L\to\infty$ and hence its distribution should
be a delta function at the
origin. Indeed the distribution in
Fig.~\ref{mcdiff} is narrow and sharply peaked at zero with a width which
deceases as $L$ increases, consistent with these expectations.

%\begin{figure}
%\includegraphics[width=\figurewidth]{fig11.eps}
%\caption{ 
%Plot of $[m_1^s m_L^s]\av^{mc}$ for different system sizes, for the
%microcanonical ensemble. Best fits to stretched exponentials with
%powers $c=0.366$ ($\chi^2=6.1$) and $c=1/3$ ($\chi^2=1380$) are also
%shown. Inset plots $\chi^2$ against $c$.}
%\label{mmdec}
%\end{figure}

Finally we look at the distributions of $C_{1L}$ and $m_1^s m_L^s$.
The scaling
variables are \cite{fisher:98} $\lambda=- \ln (C_{1L})/L^{1/2}$ and $d=- \ln
(m_1^s m_L^s)/L^{1/2}$ and relevant plots are shown in Figs.~\ref{cdist}
and \ref{mcdist} for the canonical and microcanonical ensembles respectively.
We see that in the canonical case the overall
distributions of $\l$ and $d$ are different at large arguments, but they 
match very accurately
at small values of the argument leading to the same behavior for the
averages $[C_{1L}]\av^c$ and $[m_1^s m_L^s]\av^c$
shown in Fig.~\ref{mmav}.
In the microcanonical case, the
overall distributions are quite similar but the agreement at small values of
the argument is not as good as in the canonical case. This leads to
a greater difference, shown in Fig.~\ref{mcmmav}
between the averages $[C_{1L}]\av^{mc}$
and $[m_1^s m_L^s]\av^{mc}$ than in the canonical case.

To summarize this section,  for the canonical ensemble, the end-to-end
correlation function
$[C_{1L}]\av$ falls off at criticality with a power of $L$, as
predicted analytically, Eq.~(\ref{C1L}). However, for the microcanonical
ensemble it
falls off much faster, as a stretched exponential function of distance. The
average gap, $[\Delta]\av$ falls off with a stretched exponential form at
criticality in both ensembles, with probably the \textit{same} 
dependence on $L$. 

\section{Discussion}
\label{sec:discussion}
In this paper we have looked numerically at the finite size dependence of
various quantities for the random transverse field Ising chain
(RTFIC) at criticality,
for both the canonical and microcanonical ensembles of disorder.
For quantities that span the system, $m^s$ and $C_{1L}$, finite size
scaling appears, at first glance, to indicate different correlation length
exponents for the two ensembles: $\nu=2$ for canonical and $\nu=1$ for the
microcanonical. However, in contrast to Igloi and Rieger\cite{igloi:98},
we conclude that the correct interpretation is that
the true scaling exponent is the \textit{same}
for the two ensembles, $\nu=2$, but that
the \textit{amplitude}
for the leading $\nu=2$ piece is zero, in the microcanonical
ensemble, for quantities that span the system and are thus sensitive to
the microcanonical constraint. Our reasons for this are two-fold:
\begin{enumerate}
\item
For the quantity we calculated that does not span the system, the ``bulk''
magnetization $m$, the \textit{same}
correlation length exponent $\nu=2$ was found for
both ensembles, indicating that there \textit{is} a
correlation length in the
microcanonical ensemble which diverges with the larger exponent
$\nu=2$. Hence, if this
is not seen for quantities which span the system, the explanation must be that
the amplitude is zero, not that this larger length scale does not exist.
\item
The analytical expressions for the distribution of the surface magnetization
$m^s$, first obtained by Fisher\cite{fisher:up}, show that the scaling variable
is $\bd L^{1/2}$ (canonical) and $\d L^{1/2}$ (microcanonical) demonstrating
that the true correlation length exponent is $\nu=2$ in both cases.
\end{enumerate}
Our interpretation of the data implies that the
inequality\cite{chayes:86,chayes:89}
$\nu \ge 2/d$ is
satisfied (as an equality) for the RTFIC, in contrast to the conclusion of
Pazmandi et al.\cite{pazmandi:97}.

We have also looked at the energy gap between the ground state and first
excited state at criticality. For both the canonical and microcanonical
ensembles, a stretched exponential decay describes the data. For the canonical
case, the  exponent $\mu$ (the power of $L$ in the exponential)
is exactly\cite{fisher:98} 1/3, and our numerical data are consistent with
$\mu = 1/3$ for the microcanonical case too.

The present model, the RTFIC, is integrable and thus relatively simple.
In particular, the
existence of the simple analytical expression for the surface magnetization
$m^s$, Eq.(\ref{smeq}), is surely related to 
the integrable nature of the model.
Furthermore, the microcanonical constraint
$\prod h_i = \prod J_i$ enters directly in this expression. Thus one can
plausibly see
how the constraint might affect quantities which span the system and cause
amplitudes for these quantities to vanish. However, for non-integrable
models, including models in higher dimensions, one would not expect the
microcanonical constraint to enter in a direct way even for quantities which
span the whole system. Thus it seems unlikely to us that there would be even
an \textit{apparent} difference in the critical behavior of
non-integrable models in the two ensembles.
We also note that the microcanonical and canonical ensembles of disorder have
been investigated for finite-$T$ transitions in random systems by
Aharony et al.\cite{aharony:98}
They find no difference asymptotically between the
critical behavior and finite-size effects
of the canonical and microcanonical ensembles
(which they term grand canonical and canonical respectively). 

\acknowledgments
We would like to thank D.~S.~Fisher for helpful discussions and
correspondence. This work is supported by the National Science Foundation
under grant DMR 0086287.

\appendix
\section{}
\label{appa}
Measurements made in the two ensembles are in fact related to each
other by a simple transformation at large $L$.
To see this note that
\begin{equation}
\d= \sum_{i=1}^L \xi_i ,
\end{equation}
where
\begin{equation}
\xi_i= \f{1}{L} \f{ \ln h_i - \ln J_i }
{{\rm var}(\ln h)   + {\rm var}(\ln J)}.
\end{equation}
Thus $\d$ is a sum
of $L$ uncorrelated random numbers $\xi_i,~i=1,...L$ with mean
\begin{equation}
[\xi_i]_{av}=\bd/L
\end{equation}
and variance
\begin{equation}
[\xi_i^2]_{av}-[\xi_i]_{av}^2=\f{1}{L^2}
\f{1}{{\rm var} (\ln h) +{\rm var} (\ln J)} \, .
\end{equation}
Using the central limit theorem and the definition of $\ell$ in
Eq.~(\ref{ell}) we find that,
in a canonical realization with given $\bd$
the probability, $P_{\bd}(\d)$, of obtaining the precise value $\d$ is
\begin{equation}
P_{\bd}(\d)=\left(\f{\ell}{\pi}\right)^{1/2} e^{-\ell (\d-\bd)^2}
\label{prdeq} ,
\end{equation}
for $L \to \infty$.

Now let $P^c(A,\bd)$ and $P^{mc}(A,\d)$ be the probability
distributions of some observable $A$ in the canonical and
microcanonical ensembles respectively. The two are related by 
\begin{equation}
P^c(A,\bd)= \int_{-\infty}^{\infty} P^{mc}(A,\d) P_{\bd}(\d) \, d \d .
\label{reln}
\end{equation}
Correspondingly, expectation values in the two ensembles are related by
\begin{equation}
[ A ]\av^c(\bd) =\int_{-\infty}^{\infty} [ A ]\av^{mc}(\d)\,
P_{\bd}(\d)\, d \d  .
\end{equation}

\bibliography{refs,comments}

%\begin{thebibliography}{9}
%\bibitem{fish} D.~S.~Fisher, Phys. Rev. Lett. {\bf 69}, 534 (1992);
  %Phys. Rev. B {\bf 51}, 6411 (1995). 
%\bibitem{paz} F.~Pazmandi, R.~T.~Scalettar and G.~T.~Zimanyi,
  %Phys. Rev. Lett. {\bf 79}, 5130 (1997). 
%\bibitem{chayes} J.~T.~Chayes, L.~Chayes, D.~S.~Fisher and T.~Spencer, Phys.
%Rev.\ Lett.\ {\bf 57}, 2999 (1986); J.~T.~Chayes, L.~Chayes, D.~S.~Fisher, and
%T.~Spencer, Commun.\ Math.\ Phys.\ {\bf 120}, 501 (1989).
%\bibitem{harris}A.~B.~Harris.J.\ Phys.\ C, {\bf 7}, 1671 (1974).
%\bibitem{igloi} F.~Igloi and H.~Rieger, Phys. Rev. B {\bf 57}, 11404 (1998).
%\bibitem{young} A.~P.~Young and H.~Rieger, Phys. Rev. B {\bf 53}, 8486 (1996).
%\bibitem{peschel} I.~Peschel, Phys. Rev. B {\bf 30}, 6783 (1984).
%\bibitem{fish2} D.~S.~Fisher, unpublished results.
%\bibitem{fish3} D.~S.~Fisher and A.~P.~Young, Phys. Rev. B {\bf 58},
%9131 (1998).
%\bibitem{chandra} S. Chandrasekhar, Rev. Mod. Phys. {\bf 15}, 1 (1943).
%\bibitem{aharony}A.~Aharony, A.~B.~Harris, and S.~Wiseman,
%Phys.\ Rev.\ Lett.\ {\bf 81}, 252 (1998).
%\end{thebibliography}

\end{document}